\documentstyle[12pt,fleqn,aaspp4]{article}
\newcommand{\be}{\begin{equation}}
\newcommand{\ee}{\end{equation}}

\def\max{{\rm max}}

\def\rad{{\rm rad}}
\def\obs{{\rm obs}}
\def\var{{\rm var}}

\def\kpc{{\rm kpc}}

\def\kms{{\rm km}\,{\rm s}^{-1}}
\def\r0{{R_0}}

\begin{document}

\title{Sgr A* ``Visual Binaries'': A Direct Measurement of the Galactocentric Distance}

\author{Samir Salim}
\affil{E-mail: samir@astronomy.ohio-state.edu}
\author{Andrew Gould}
\affil{E-mail: gould@astronomy.ohio-state.edu}
\affil{Ohio State University, Department of Astronomy, 
174 West 18th Ave., Columbus, OH 43210}

\begin{abstract}
	We present a new geometrical method for measuring the distance to the Galactic center ($\r0$) by solving for the Keplerian orbit of individual stars bound to the black hole associated with the Sgr A* from radial velocity and proper motion measurements. We identify three stars to which the method may be applied, and show that 1-5\%  accuracy of $\r0$ can be expected after 15 years of observing, and 0.5-2\% after 30 years of observing, depending on what the orbital parameters of these three stars turn out to be. Combining the measurements of the three stars with favorable orbital parameters leads to even more precise values. In the example that we present, such combined solution yields 4\% accuracy already by the year 2002. All these estimates assume that annual position measurements  will continue to be made with the $\sigma_\rho = 2$ mas precision recently reported by Ghez et al.\ The precision of the distance measurement is relatively insensitive to the radial velocity errors, provided that the latter are less than $50\, \kms$. Besides potentially giving an estimate of $\r0$ that is better than any currently in use, the greatest advantage of this method is that it is free from systematic errors.

\end{abstract}
\clearpage

\section{Introduction}

	The distance to the center of the Galaxy ($\r0$) is to Galactic astronomy what the Hubble constant ($H_0$) is to extragalacatic astronomy and cosmology. Many of the parameters we determine for the Galaxy depend on $\r0$. Fixing its value will directly give us a precise value of circular velocity $\Theta_0$ from the measured proper motion of Sgr A*. The circular velocity and distance in turn enter in estimates of the mass of the Galaxy. Having a precise value of $\r0$ would allow us to recalibrate distance indicators such as RR Lyraes, and compare them with local RR Lyraes. This would give us an idea of applicability of the local RR Lyrae calibration to globular clusters and the LMC. This can to some extent also be done for the Cepheids, which are an important step in the extragalactic distance ladder. 

	As with the Hubble constant, the distance to the center of the Galaxy is still debated (although with perhaps less controversy), 70 years after it was first estimated by Shapley (1918). In 1993, Reid presented an overview of then current methods and suggested a weighted average of $\r0 = 8.0 \pm 0.5$ \kpc. This is a convenient value which is still used by many researchers (although the IAU maintains its recommended value at $\r0 = 8.5$ \kpc). In the closing sentence of his review, Reid (1993) expects $\r0$ to be known with 3\% uncertainty by the end of the decade. How close are we to this goal today? In the period after the Reid's (1993) article there were four new estimates of $\r0$, one of which was based on a new method. 

	The first new estimate uses RR Lyrae observations in the near-infrared (IR). Apparent magnitudes suffer less interstellar extinction in the near-IR than in the optical. Smaller brightness amplitudes in the near-IR also permit an estimate of the mean magnitude with fewer measurements. Using 58 RR Lyraes in Baade's Window, Carney et al.\ (1995) report $\r0 = 7.8 \pm 0.4$ \kpc\, which is consistent with Reid's value and has 5\% uncertainty. However, the real uncertainties turn out to be greater, as this result depends on the distance-scale model used. In fact, the authors report two possible results that differ by 14\%. 
	Metzger, Caldwell \& Schechter (1998) measured $\r0$ based on Cepheids and a model of the Galactic rotation. They find $\r0 = 7.66 \pm 0.32$ \kpc. The 4\% uncertainty is a combination of many different systematic errors (Cepheid calibration, reddening, etc) that enter in the final result. The value of $\r0$ itself is also model dependent, and an additional uncertainty comes from an unknown component of the ellipticity of the stellar orbits.

	The Cepheid distance scale was recalibrated using the {\it Hipparcos} satellite. This led to a new estimate of $\r0$ by Feast \& Whitelock (1998). They report a ``long'' scale of  $\r0 = 8.5 \pm 0.5\, \kpc$, i.e., a 6\% uncertainty.
	
	{\it Hipparcos} data also facilitated the introduction of a new method using red clump stars. Paczy\'nski \& Stanek (1998) used {\it Hipparcos} parallaxes to calibrate nearby red clump stars and applied this calibration to find the distances of the red clump stars near the Galactic center. Due to the great number of these stars in Baade's Window, the formal uncertainty of $\r0$ is very small. However, various systematics produce a final estimate of $\r0 = 8.4 \pm 0.4$ \kpc, i.e. 5\% accuracy. Later, this result was corrected to $\r0 = 8.2 \pm 0.15 \pm 0.15$ \kpc \, (Stanek \& Garnavich 1998).
 
	As we can see, there is still no agreement on a single value of $\r0$, and all of the estimates have significant uncertainties. Notably, none of these methods falls into a category of primary methods -- those that are not using a ``standard candle'' or a Galactic rotation model (Reid 1993). One would prefer to use a direct method in order to avoid systematics. Even the $\rm{H_2O}$ maser method (Genzel et al.\ 1981; Reid 1993) is model dependent. Also, the accuracy of this method seems to be limited to about 15\%.

	An attractive way to directly determine $\r0$ is to measure the trigonometric parallax of Sgr A* using VLBI. However, significant technical obstacles remain for this approach (Reid 1996, private communication). The {\it Space Interferometry Mission (SIM)} will be sensitive enough to measure parallaxes with 3\% precision at the Galactocentric distance, but since it operates in the optical part of the spectrum, {\it SIM} cannot look directly at the Galactic center.

	Here we suggest a new, purely geometric method, which became feasible with the recent advances in speckle imaging of the stars surrounding the Galactic center (Eckart \& Genzel 1997; Genzel et al.\ 1997; Ghez et al.\ 1998). In essence, it is the classic {\it visual binary} method which has been used for decades to measure the masses of and distances to binary stars. We apply this method to stars orbiting around the massive object in the center of the Galaxy (presumably a black hole). This method requires observations of the proper motions and radial velocities of several stars near Sgr A* during the next decade or so, in order to constrain the shape and the physical size of the orbit. These, together with the angular size of the orbit will in turn yield $\r0$. This is not a statistical method: each star individually leads to a value of $\r0$ whose uncertainties depend only on the precision of the proper motions and radial velocity measurements. Its basic advantages are that it is free of systematic errors and that it can possibly measure $\r0$ accurate to 1-5\% after 15 years of observing, and 0.5-2\% after 30 years of observing.

	The orbital solution to a visual binary automatically gives the
heliocentric radial velocity of the center of mass which in this case is 
nearly identical to Sgr A*.  Since the velocity of the Local Standard of
Rest (LSR) is known, the radial velocity of Sgr A* immediately gives the
velocity of the LSR relative to Sgr A*.  If this velocity is found
to differ significantly from zero, it can only mean that either Sgr A*
is not at rest with respect to the Galaxy or that the LSR is moving
radially with respect to the Galaxy.  Gould \& Popowski (1998) find that
the LSR is moving radially outward relative to local halo stars at
$-4.0\pm 8.5\,\kms$.  Measurement of the velocity of Sgr A* would therefore
allow one to distinguish between these two possibilities.

\section{Physical Principles}

	Eckart \& Genzel (1997) and Ghez et al.\ (1998) have measured the proper motions of many stars within $1''$ of Sgr A*. These proper motions are actually Keplerian {\it orbital} motion of stars around the central mass of $M \cong 2.6 \times 10^6 M_{\odot}$ (this assumption will be discussed in more detail in \S\ 5.). As yet, the time baseline of these observations ($\sim 2$ years) is not sufficient to show these proper motions as making an arc around Sgr A*. Let us consider star S0-1 from the Ghez et al.\ (1998) catalogue (S1 in Genzel et al.\ 1997 catalog), at a separation of $0.\hskip-2pt''1$ from Sgr A*, the closest so far detected. Of all the stars in the vicinity of Sgr A*, S0-1 is the most likely to show the signs of orbital motion in near-term future observations. How long do we have to wait until this happens? If we assume a circular orbit, and that the projected separation at which we now see S0-1 is close to the angular radius of its orbit, then Kepler's law gives us a period of 17 years. This is just a rough estimate, but it gives us an idea that the full orbit of S0-1 can be traced in a reasonable amount of time.

	What will learning about the orbit of S0-1 tell us about $\r0$? For simplicity, assume that S0-1 is in a circular orbit of period $P=15$ yr, with a major-axis $a = \r0 \alpha$, where $\alpha$ is the current separation of $\alpha=100$ mas. Let the angle of inclination be $i$, so that the projected semi-minor axis is $b=a \cos i$. The observed dispersion in radial velocity measurements is related to these quantities by
\be
[\var (v_{\rad})]^{1/2} = (\langle v_{\rad}^2 \rangle - \langle v_{\rad} \rangle^2 )^{1/2} = {2 \pi a \sin i \over P} = {2 \pi \alpha \sin i \over P}\, \r0
\ee
	Let us for the moment assume that $P$ and $\var (v_{\rad})$ can be measured with perfect accuracy. Then the fractional error in the determination of $\r0$ is equal to the fractional error in the measurements of $\alpha \sin i$. The fractional error in $\alpha$ is simply
\be
{\delta \alpha \over \alpha} = {\sigma_\rho \over \alpha} \sqrt{2\, \rm yr \over P} = 0.007\, \left({\sigma_\rho \over 2 \rm mas}\right) \left({\alpha \over 100 \rm mas}\right)^{-1} \left({P \over 15 \rm yr}\right)^{-1/2} 
\ee
implying that the method is potentially very precise. However, for low-inclination orbits, the fractional error in $\alpha \sin i$ is much greater than in $\alpha$. Specifically
\be
{\delta(\alpha \sin i) \over \alpha \sin i} = {\delta(\sqrt{\alpha^2-\beta^2}) \over \sqrt{\alpha^2-\beta^2}} =  {\sqrt{\alpha^2+\beta^2} \over \alpha^2-\beta^2}\, \delta \alpha = {\sqrt{1+\cos^2 i} \over \sin^2 i}\, {\delta \alpha \over \alpha},
\ee
where $\beta = b / \r0$. That is, the fractional error in $\r0$ is larger than the fractional error in $\alpha$ by a factor of $2.5$ for $i=45^{\circ}$, and by a factor of $5.3$ for $i=30^{\circ}$. The fractional error in the period will be $\delta P / P \sim \delta \alpha / \alpha$, so it is appropriate to ignore it in this simplified analysis.

	How good do the radial velocity measurements have to be so that they do not dominate the error in the determination of $\r0$? The fractional error in the dispersion measurement must be smaller than the fractional error in $\alpha \sin i$. That is
\be
{\delta \sqrt{\var (v_{\rad})} \over \sqrt{\var (v_{\rad})}} = {\sigma_{v_{\rad}}(P/2\, \rm yr)^{-1/2} \over v_0 \sin i} < {\delta(\alpha \sin i) \over \alpha \sin i} = {\sqrt{1+\cos^2 i} \over \sin^2 i}\, {\delta \alpha \over \alpha},
\ee
where $v_0 \equiv 2\pi a/P$, and where we have assumed that there are annual radial velocity measurements with accuracy $\sigma_{v_{\rad}}$. Equation (4) can be expressed
\be
\sigma_{v_{\rad}} < {\sigma_{\rho} \over \alpha}\, {\sqrt{1+\cos^2 i} \over \sin i}\, v_0 = 56\, \kms \left({\sigma_\rho / \alpha \over 0.02}\right) {\sqrt{\csc^2 i + \cot^2 i} \over 2}\, \left({v_0 \over 1400\, \kms}\right).
\ee
Hence, the velocity errors will not dominate unless they are greater than $50\, \kms$.

	This simple analysis for the case of a circular orbit demonstrates the potency of the method. In the next section we will do the exact (although intuitively less clear) treatment of this problem in the case of orbits with arbitrary eccentricity and orientation. We present the accuracies obtainable for $\r0$ from the three stars closest to Sgr A*. In \S\ 5., we discuss the validity of our basic assumptions and some specific details.

\section{The Method}

	The distance to Sgr A*, $\r0$, is one of the 13 parameters required to describe the Star-Sgr A* binary orbit, and the precision of its measurements depends not only on the observational errors, but on the actual values of the 13 parameters, i.e. on the specific orbit. We can only specify the precision of this distance measurement as a function of these 13 parameters. At first sight, this appears unwieldy. However, as we now show, 11 of the parameters are known sufficiently well that the uncertainty in $\r0$ does not depend on them in a major way. They can therefore be considered fixed, and it is only necessary to specify the uncertainty in $\r0$ as a function of the two remaining free parameters.

	The 13 parameters can be taken to be: (1) the six phase-space co-ordinates of Sgr A*, (2) the six phase-space co-ordinates of the orbiting star relative to Sgr A*, and (3) the mass $M$ of Sgr A*. The six phase-space co-ordinates of Sgr A* and the mass of Sgr A* can be considered known. For example, we take the distance to be 8 kpc. However, if the distance is 5\% higher, 8.4 kpc, our estimate of the distance {\it error} will be off by a similar amount, which is quite small compared to the factor $\sim 5$ range of values we will explore for the distance error. Similarly, we take the mass of Sgr A* to be $M \cong 2.6 \times 10^6 M_{\odot}$, but if it turns out to be 5\% higher or lower, our estimate of the distance error will have to be rescaled by a similar amount. Nothing in the error analysis depends in any way on the values of the two-dimensional angular position, the two-dimensional angular velocity and the radial velocity of Sgr A*, so these quantities can be truly ignored.

	Four quantities are measured: the two-dimensional angular position and velocity of the star. Unfortunately, these measurements are relative to the frame of the near-IR speckle images, not Sgr A*. However, Menten et al.\ (1997) have aligned this IR frame with the radio frame in which Sgr A*'s position and proper motion are measured. Using this alignment, Ghez et al. (1998) fix the position of Sgr A* to $\pm 10$ mas. Ultimately, the visual-binary orbital solution will fix this position with an order of magnitude better precision.

	For the orbit to be completely specified, two additional parameters are required. For convenience, we take these two parameters to be the period $P$ and inclination $i$. Thus, for each $P$ and $i$ we construct an orbit that is consistent with the observable quantities: the two-dimensional angular position and two-dimensional angular velocity. Not all values in the $P-i$ parameter space produce an orbit. When they do there are four possible solutions. There are two sets of two solutions that each comprise a pair of orbits that are symmetrical  in the plane of the sky and so have the same error structure. Thus, there are effectively two different solutions (two different orbits) for each $P$ and $i$. We calculate orbital parameters for all allowed orbits having periods $P<40$ yr (we choose this as an upper limit in our study since longer periods will take too long to constrain the orbit observationally). 

	After we find the parameters for all the possible orbits of a given star we simulate a set of observations made once a year. Observations consist of measuring the position and the radial velocity. We introduce the radial velocity measurements in the fifth year of observations (i.e. in the year 2000, since we set $t_{\obs}=0$ at the epoch of the first Ghez et al.\ (1998) observations, i.e., the year 1995), when we expect adaptive-optics spectroscopy to become good enough to produce these data. We assume the current positional measurement accuracy (2 mas) throughout, and an error in the radial velocity measurements of $\sigma_{v_{\rad}} = 20\, \kms$. The choice of $\sigma_{v_{\rad}}$ will be discussed in the next section.

	We determine the evolution of the accuracy of $\r0$ as the observations accumulate by calculating the $13 \times 13$ covariance matrix for the $n=13$ parameters of the model. To do so, we first write the three observables $F^m(t)$ ($m = 1,2,3$ for $F^m= \rho_x, \rho_y, v_{\rad}$) as functions of the time $t$ and of the 13 parameters which we label $a_i$, $F^m(t; a_1, \ldots, a_{13})$. We then linearize $F^m$ in the neighborhood of the solutions, $a_i^\ast$.
\be
F^m_{lin}(t) = \sum^{n}_{i=1} (a_i-a_i^\ast) f_i(t) + F^m(t; a_1^\ast, \ldots, a_{13}^\ast)
\ee
where
\be
f^m_i(t) \equiv {\partial F^m \over \partial a_i}.
\ee
We then specify the solution using seven classical parameters for the orbit of the star relative to Sgr A*, and six to specify the three-dimensional position and velocity of Sgr A* itself. Specifically, we have
\be
F^m(t) = G^m(t) + \rho_{m,0} + \mu_m t, \qquad (m=1,2)
\ee
\be
F^3(t) = \r0\, {\partial G^3(t) \over \partial t} + v_{r,0}
\ee
where $(\rho_1, \rho_2)_0$ is the initial position of Sgr A* and $(\mu_1, \mu_2)$ is its proper motion relative to the IR star frame, $v_{r,0}$ is the heliocentric radial velocity of Sgr A*, and $G^m(t)$ is the three-dimensional orbit of the star relative to Sgr A*, in angular units. That is,
\be
G^m(t) = \sum_{pqr} {\cal R}_{mp}^1 (t_1) {\cal R}_{pq}^2 (t_2) {\cal R}_{qr}^3 (t_3) H^r(t),
\ee
\be
H^1(t) = \alpha (\cos \psi - e), \qquad H^2(t) = \alpha (1-e^2)^{1/2}\sin \psi, \qquad H^3(t)=0,
\ee
\be
\psi - e \sin \psi = \omega t + \phi.
\ee
Here, $\alpha$ is the angular semi-major axis, $e$ is the eccentricity, $\omega = 2 \pi / P$, $\phi$ is the phase of the orbit at $t=0$, and ${\cal R}^1$, ${\cal R}^2$, and ${\cal R}^3$ are the rotations which we take to be around the 3, 1, and 3 axes respectively.

	We then create the matrices $b$, and $c$ with elements
\be
b_{ij}= \sum_{k,m}\, {\partial F^m \over \partial a_i}\, {\partial F^m \over \partial a_j}\, {1 \over (\sigma^m_k)^2}, \qquad c = b^{-1}
\ee
where $\sigma^m_k$ is the error in the $k$-th measurement of $F^m(t_k)$. The index $k$ steps through the ``data'' -- in our case the observations made each year. The square of fractional uncertainty of the parameter $a_j$ is just the covariance matrix element $c_{jj}$. One of these parameters is $\r0$. This allows us to predict the accuracy of finding $\r0$ after $t_{\obs}$ years of observing. In the next section we present the results of these calculations for the three stars closest to Sgr A*.

\section{Predictions of the Uncertainty of $\r0$}

	How does the accuracy of the $\r0$ determination evolve as the observations accumulate at the rate of one per year? Let us illustrate this using the star S0-1, presently the closest to Sgr A*. For each $P$ and $i$ we will for now consider only one of the two possible orbits. 

	In Figure 1 we fix the inclination at $i=45^{\circ}$ and explore all the possible orbits with this inclination. Different lines correspond to different possible periods in integer increments. The short-dashed line corresponds to the shortest possible period of $P=20$ yr, while the long-dashed one represents the orbit of $P=40$ yr, the upper limit that we explore. Each line represents the evolution of the fractional uncertainty of $\r0$, as the observing time $t_{\obs}$ progresses. The first thing to notice is that for the orbits of a fixed inclination, $\sigma_{\r0}/\r0$ will not depend strongly on the period. The plot shows that 5\% accuracy is achieved after about 12 years of observing, 2\% after some 20 years, and 1\% accuracy after approximately 30 years. After this time, which corresponds to observing a star throughout its entire orbit, a plateau is reached, i.e. additional observations reduce the uncertainty of $\r0$ only as $t_{\obs}^{-1/2}$.  

	We now explore the set of possible orbits with $i=55^{\circ}$. This is close to the highest possible inclination of $i_{\max}=57^{\circ}$ that star S0-1 could have. The evolution of $\sigma_{\r0}/\r0$ is shown in Figure 2. The shortest possible orbit (short-dashed line) now has $P=32$ yr. Compared to $i=45^{\circ}$ case, the uncertainties of $\r0$ now depend even more weakly on period. The trend of $\sigma_{\r0}/\r0$ evolution is also different. An accuracy of 5\% is reached after 16 years, i.e. later than in the case of $i=45$, while 2\% and 1\% accuracies are obtained after about the same time as for $i=45^{\circ}$, 20 and 30 years, respectively. While in the case of $i=45^{\circ}$ orbits, the maximum accuracy is reached at 0.7\% level, now it is much lower, 0.4\%. This illustrates that this method can potentially yield a very low uncertainty for $\r0$: it all depends on what the orbit of this star (and the other two) turn out to be. It should be noted that the second set of solutions (for a given $P$ and $i$) essentially give the same trends of $\sigma_{\r0}/\r0$ evolution. 

	As we mentioned above, the radial-velocity errors are assumed to be $\sigma_{v_{\rad}} = 20\, \kms$. Since these measurements have not yet been made, we do not know what their precision is going to be. However, we can explore how sensitive the determination of $\r0$ is to a range of different values of $\sigma_{v_{\rad}}$. In Figure 3 we show the evolution of $\sigma_{\r0}/\r0$ for S0-1 for the $P=30$ yr, $i=45^{\circ}$ orbit (one of the lines in Figure 1) for various assumed values $\sigma_{v_{\rad}} = 5, 10, 20, 50$ and $100\, \kms$. The basic result is that $\sigma_{\r0}/\r0$ is not strongly affected by $\sigma_{v_{\rad}}$ over this range. At late times ($t_{\obs}>30$ yr) $\sigma_{\r0}/\r0$ depends only very weakly on $\sigma_{v_{\rad}}$. In the region of $t_{\obs}$ where this dependence is strongest we see that $\sigma_{v_{\rad}} = 20\,\kms$ (thick line) is the highest value that is still close to the lines of low $\sigma_{v_{\rad}}$. That is why we adopt $\sigma_{v_{\rad}} = 20\,\kms$ throughout this paper. The $\sigma_{\r0}/\r0$ dependence on $\sigma_{v_{\rad}}$ is much less for the orbits with short periods (regardless of inclination), and only somewhat greater for longer periods.

	Finally, we move on to exploring the entire $P-i$ parameter space. We do so by making contour plots for S0-1 in the $P-i$ plane of $\sigma_{\r0}/\r0$ after some fixed observing time. (We use contours showing 10, 5, 2, 1, and 0.5\% error). We examine two cases: $t_{\obs}=15$ yr and $t_{\obs}=30$ yr. The outermost contour in all of the contour plots actually limits the parameter space of possible orbits. Figure 4 displays contours of $\sigma_{\r0}/\r0$ for orbits with different $P$ and $i$ after $t_{\obs}=15$ yr. It shows that orbits of higher inclination allow more precise determination of $\r0$. Since for each $P$ and $i$ we have two possible orbits, in Figure 5 we show the contours for the second set of solutions, also after $t_{\obs}=15$ yr. The region where $\sigma_{\r0}/\r0 < 5\%$ is smaller in this case. After $t_{\obs}=30$ yr, the two solutions (Figs. 6,7) are much more alike. The contours of high $\r0$ precision (better than 1\%) are again in the high inclination range and towards shorter periods. This corresponds to short orbits (full orbit is traced faster) that are away from face-on orientation where the fractional error in $\alpha \sin i$ is high. See \S\ 2.

	Star S0-1 is the closest star from Sgr A* ($0.\hskip-2pt''114$) found to date and has the highest measured two-dimensional velocity ($\sim 1400\, \kms$)(Ghez et al.\ 1998). Therefore, it represents the first logical choice for the analysis we showed. However, we find two other stars in the Ghez et el. (1998) catalog that can also be used for finding $\r0$ in $t_{\obs}=30$ yr, and one of them produces useful results in $t_{\obs}=15$ yr. These are S0-2 ($0.\hskip-2pt''151$ from Sgr A*) and S0-3 ($0.\hskip-2pt''218$ from Sgr A*). They are also known as S2 and S4 in the Genzel et al.\ 1997 catalog. (S3 from Genzel et al.\ 1997 is another candidate, but its detection is  ambiguous in the Ghez et al.\ 1998 catalog which we use throughout this paper.

	We find (but do not show) that the $P-i$ parameter space for the star S0-2 looks substantially different from S0-1. Highly inclined orbits (almost edge-on) are allowed for fairly short orbital periods. After $t_{\obs}=15$ yr some orbits already produce $\r0$ estimates with better that 1\% errors. This is several times better than in the case of S0-1. After $t_{\obs}=30$ yr, some orbits of S0-2 even give a better than 0.5\% estimate of $\r0$. Finally, for the star S0-3, after $t_{\obs}=30$ yr some orbits yield a better than 1\% estimate of $\r0$. No useful estimate of $\r0$ can be obtained from S0-3 after $t_{\obs}=15$ yr. This is not surprising since the shortest possible orbits have $P=19$ yr.

	How much would the three stars together, compared to each of them individually, contribute to a more certain value of $\r0$? One can simply take a weighted average of the estimates of $\r0$ from each single star. To illustrate this, in Figure 8 (note the extent of {\it y}-axis compared to Figs. 1-3) we choose some values for the orbital parameters for each of the stars S0-1 ($P=30$ yr, $i=45^{\circ}$), S0-2 ($P=20$ yr, $i=70^{\circ}$) and S0-3 ($P=30$ yr, $i=75^{\circ}$). The evolution of $\sigma_{\r0}/\r0$ from each of them is shown by thin solid lines. The uncertainty obtained by averaging is shown by a thick dashed line. However, observing the three stars, even for a short time, can produce an
even better estimate of $\r0$ than the simple weighted average. This is because partial orbits of more than one star can together constrain the position, velocity and mass of Sgr A* much better than each orbit individual: each star must orbit a common mass at a common center. We show the reduced uncertainty by the solid thick line. In the first few years this combined solution produces a very significant improvement compared to the naive statistical treatment.  Afterwards, it remains better by some 40\%. Note that in this example, 4\% precision can be obtained already in the eighth year of observation, and 1\% precision only two years after that, i.e. by 2003 and 2005, respectively. Ultimately, 0.2\% accuracy of $\r0$ could be achieved. We also performed the same analysis excluding the observations of S0-3. We find that S0-3 contributes significantly (mostly by constraing the mass of Sgr A*) even at times when the observations of S0-3 alone do not tell us much about $\r0$. This indicates that perhaps even the stars farther away, that alone cannot determine $\r0$ (those with $P>40$ yr), might contribute by constraining the mass of Sgr A*. Therefore, it would be advisable to keep track of other stars near Sgr A* in addition to the three closest.

	We note that in this example, the radial velocity of Sgr A* is
measured to $5\,\kms$ precision after 20 years and $3\,\kms$ after 30 years,
roughly half the measurement errors of Gould \&
Popowski (1998).  See \S\ 1.

\section{Discussion and Conclusions}

	Besides the possibility of giving us a precise measurement of $\r0$, this method is very powerful because it does not rely on some model or a calibration. It has a single underlying assumption: stars in the vicinity of Sgr A* move in a Keplerian potential. We believe that this is a robust assumption which is consistent with the available observational data. Nevertheless, it is possible in principle that in addition to the point mass at the center of the potential there is a continuous mass distribution composed of many faint stars. This would cause the orbits to precess. However, this motion, if significant, would show itself after the orbit has been fully traced, and can then easily be accounted for. We would still obtain a precise estimate of $\r0$.

	Recently, Munyaneza, Tsiklauri \& Viollier (1998) have explored an exotic version of this scenario with an extended object at the Galactic center composed of degenerate, self-gravitating heavy neutrino matter. They show that such an object would alter the orbit of S0-1 so as to make it distinctly different from a Keplerian orbit, and that positional measurements alone will be able to resolve this over the next 10 years. Distinguishing between the black-hole and extended-object scenario will become even easier once the radial velocity curve of S0-1 is measured. 

	In \S\ 2 we argued that in the case of circular orbits, only relatively crude radial velocity measurements ($\sigma_{v_{\rad}} \leq 50\, \kms$) are required. We now investigate the role of radial velocity measurements more closely. In Figure 3 we show the fractional error in $\r0$ as a function of time for S0-1 assuming $i=45^{\circ}$ and $P=30$ yr, and for various values of $\sigma_{v_{\rad}}$. Also shown in this plot (dashed curve) is the fractional error in the quantity $\alpha \sin i / P$. This would be exactly equal to $\sigma_{\r0}/\r0$, if there were perfect knowledge of $\var(v_{\rad})$ (i.e. perfect velocity measurements) but none of the radial-velocity information were used to disentangle degeneracies among the orbital prameters.

	At late times ($t_{\obs}>P$) all the curves converge with the dashed curve lying slightly below the others. This shows that high-precision velocity measurements are indeed of small value if the observations are carried out for at least one period. However, at early times, the dashed curve lies {\it well above} the others. This shows that for observations shorter than the period, the radial velocity measurements play a significant role in disentangling the degeneracies among the orbital parameters, and hence can significantly help determine the distance $\r0$. 

	The success of the method we present, in terms of precision, will eventually depend on how lucky we are to see the stars in the vicinity of Sgr A* in favorable orbits. In any event, this method will produce an estimate of the distance to the Galactic center that is free from the various systematic effects that affect other methods currently in use.

{\bf Acknowledgements}:  
This work was supported in part by grant AST 97-27520 from the NSF.

\clearpage

\begin{figure}
\caption[junk]{\label{fig:one}
Evolution of $\r0$ fractional uncertainty from S0-1 orbits with $i=45^{\circ}$ ($t_{\obs}$ is in years). Different lines correspond to different integer orbital periods. The short-dashed line corresponds to the minimum possible period of $P=20$ yr, while the long-dashed line represents the orbit of $P=40$ yr, our upper limit.
}
\end{figure}

\begin{figure}
\caption[junk]{\label{fig:two}
Evolution of $\r0$ fractional uncertainty from S0-1 orbits with $i=55^{\circ}$ ($t_{\obs}$ is in years). Different lines correspond to different integer orbital periods. The short-dashed line corresponds to the minimum possible period of $P=32$ yr, while the long-dashed line represents the orbit of $P=40$ yr, our upper limit.
}
\end{figure}

\begin{figure}
\caption[junk]{\label{fig:three}
Evolution of $\r0$ fractional uncertainty from S0-1 orbit ($P=30$ yr, $i=45^{\circ}$, $t_{\obs}$ is in years) with different assumptions for the measured $\sigma_{v_{\rad}} = 5, 10, 20, 50$ and $100\, \kms$ (upper solid line corresponds to $100\, \kms$). Bold line represents $\sigma_{v_{\rad}} = 20 \kms$. Dashed line is the fractional error in $\alpha \sin i\, /P$. See \S\ 5.
}
\end{figure}

\begin{figure}
\caption[junk]{\label{fig:four}
Contour plot of $\r0$ fractional uncertainty from S0-1 orbit, after $t_{\obs}=15$ yr. Solution 1.
}
\end{figure}

\begin{figure}
\caption[junk]{\label{fig:five}
Contour plot of $\r0$ fractional uncertainty from S0-1 orbit, after $t_{\obs}=15$ yr. Solution 2.
}
\end{figure}

\begin{figure}
\caption[junk]{\label{fig:six}
Contour plot of $\r0$ fractional uncertainty from S0-1 orbit, after $t_{\obs}=30$ yr. Solution 1.
}
\end{figure}

\begin{figure}
\caption[junk]{\label{fig:seven}
Contour plot of $\r0$ fractional uncertainty from S0-1 orbit, after $t_{\obs}=30$ yr. Solution 2.
}
\end{figure}

\begin{figure}
\caption[junk]{\label{fig:eight}
Evolution of $\r0$ fractional uncertainty from each of the stars ($P_1=30$ yr, $i_1=45^{\circ}$; $P_2=20$ yr, $i_2=70^{\circ}$; $P_3=30$ yr, $i_3=75^{\circ}$) separately (thin solid lines), and when combined in a straightforward statistical way (thick dashed line), and a combined solution with a common position, velocity and mass of Sgr A* (thick solid line).
}
\end{figure}

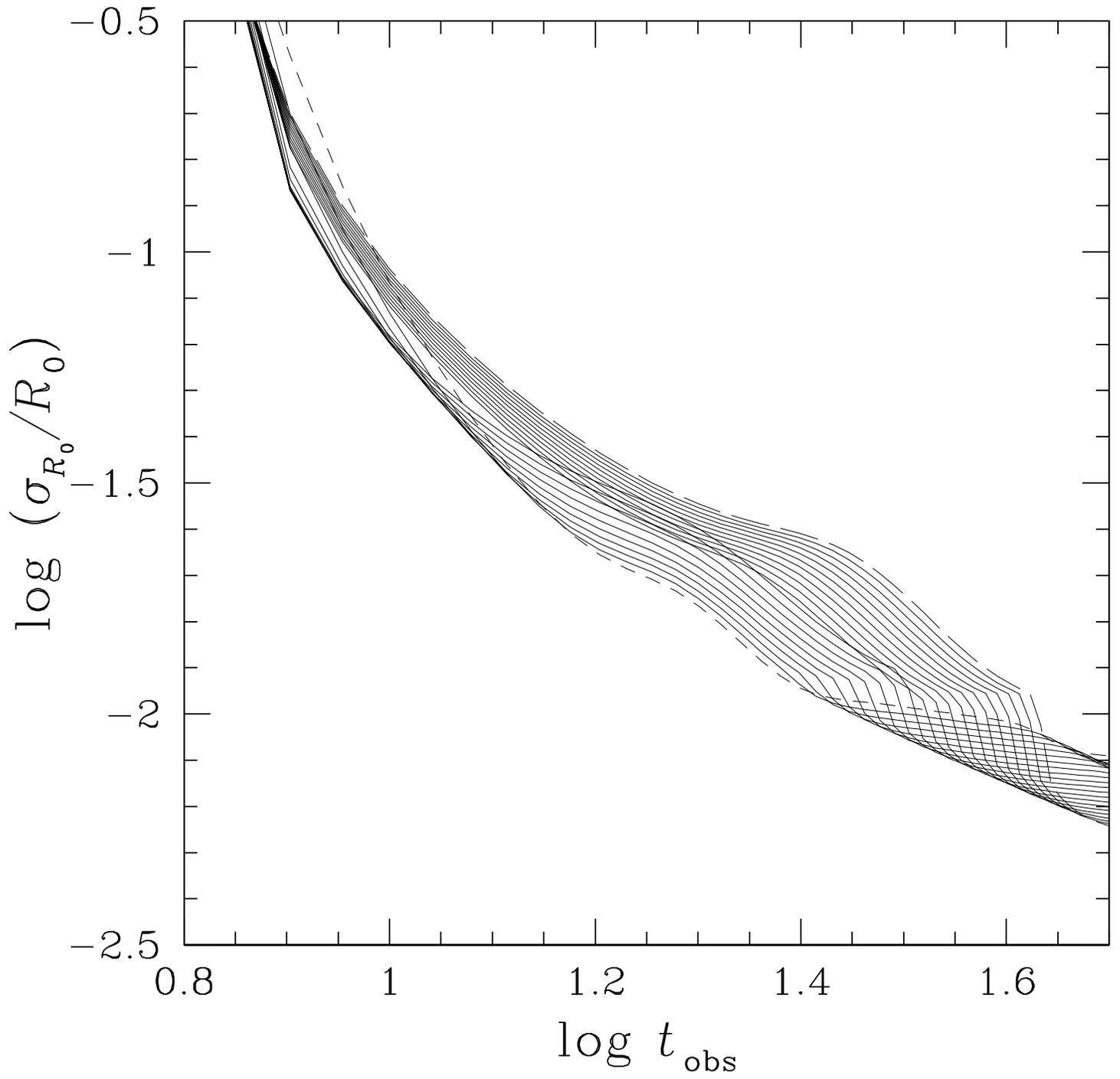
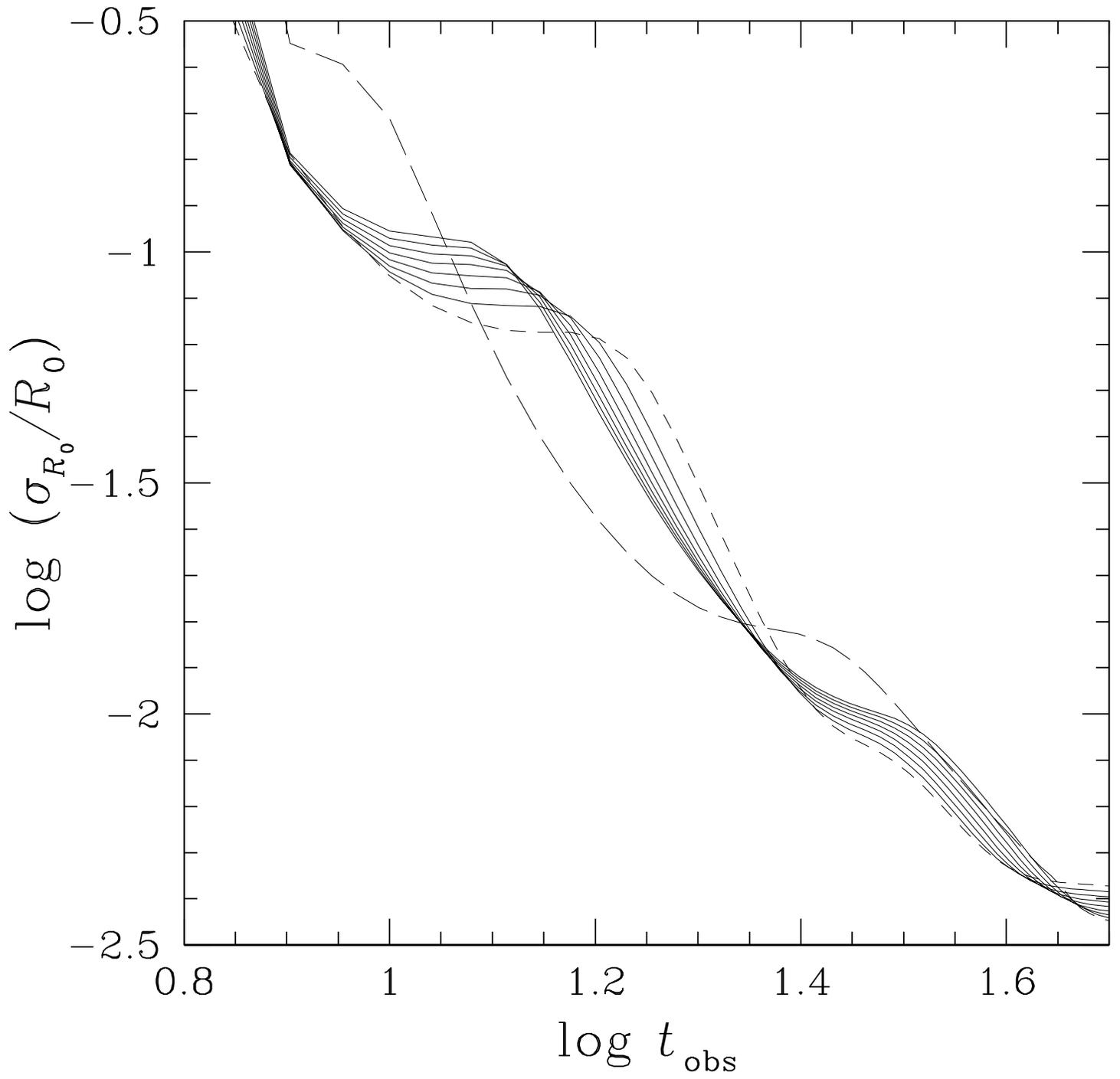
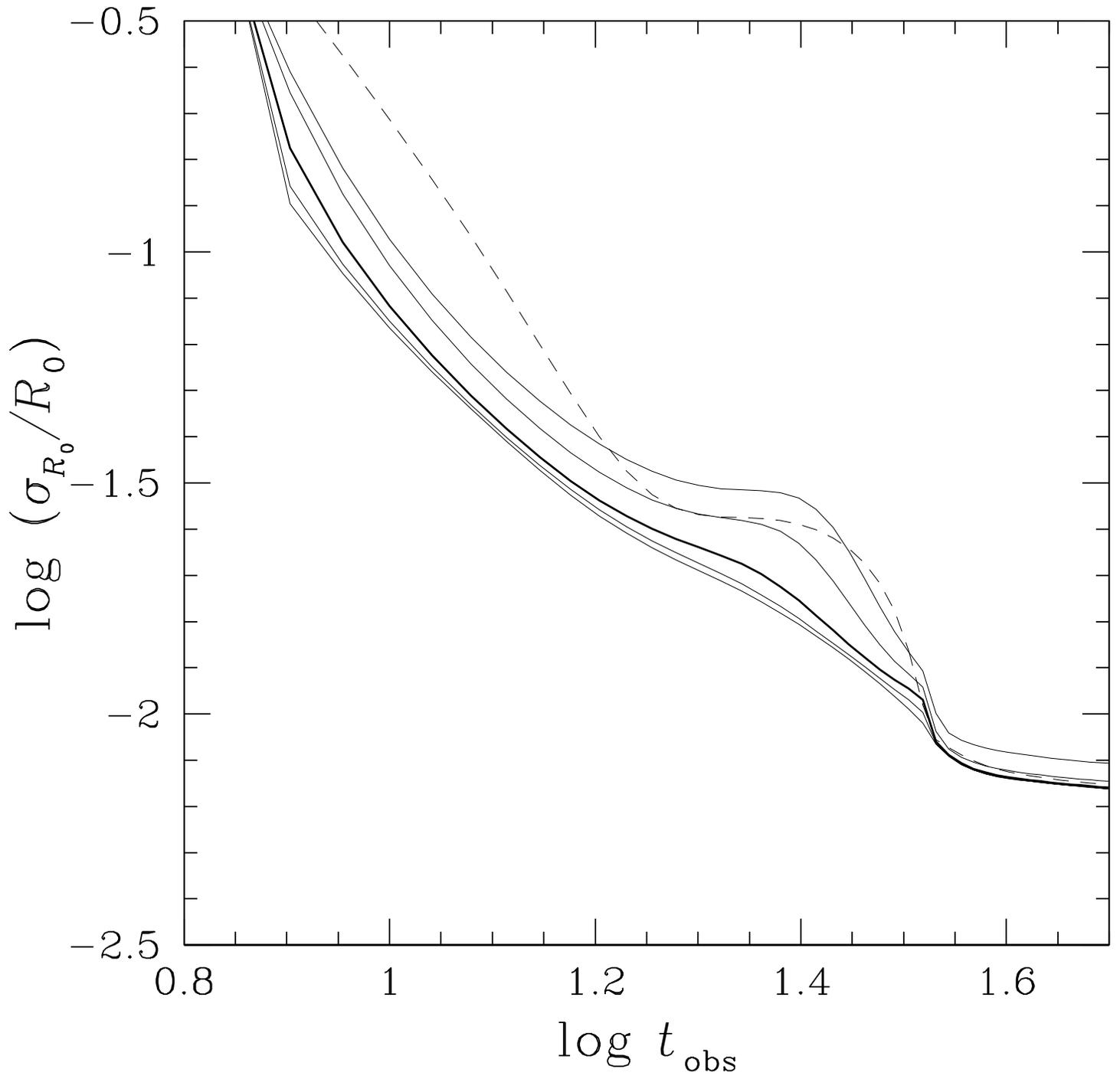
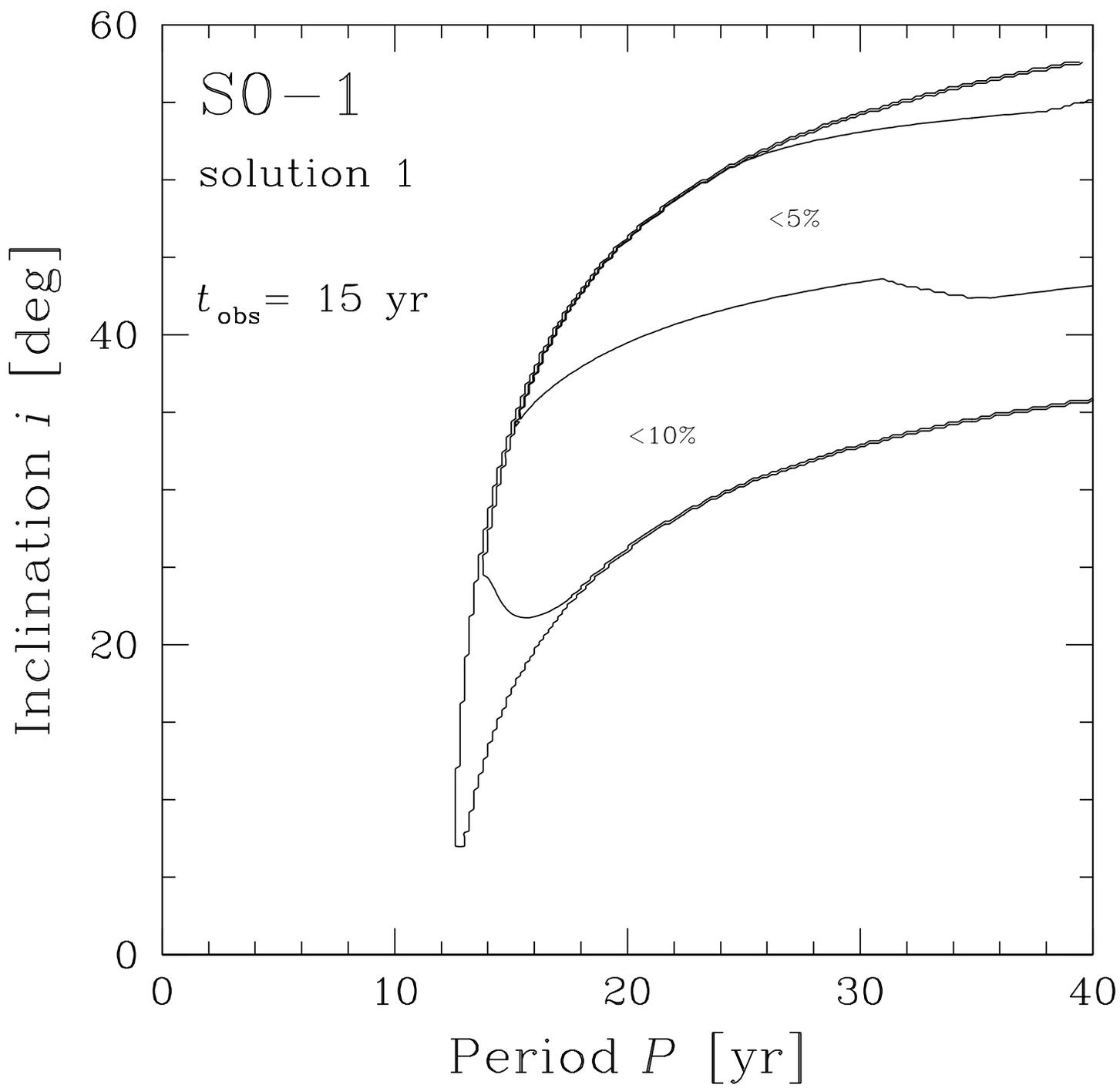
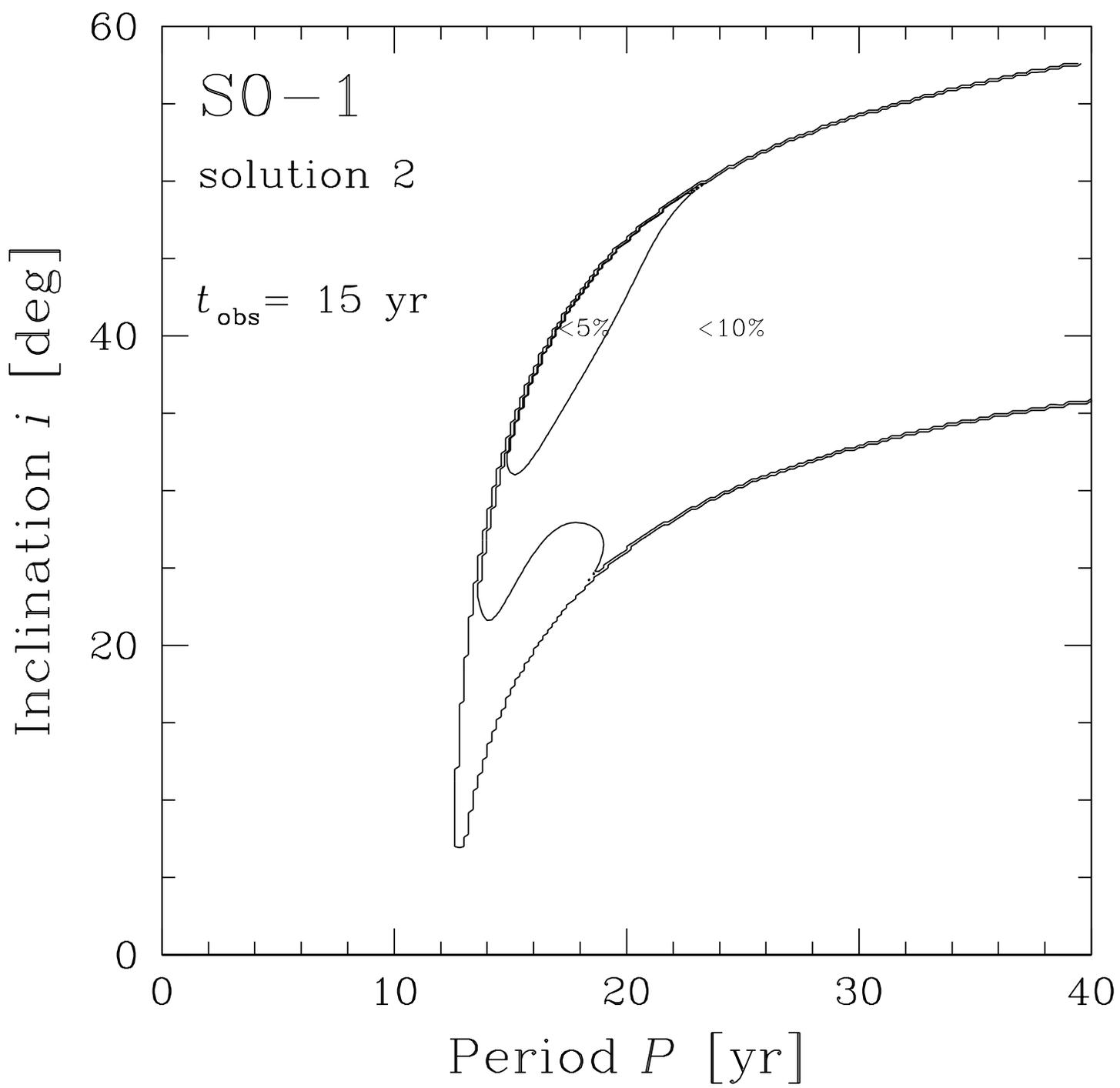
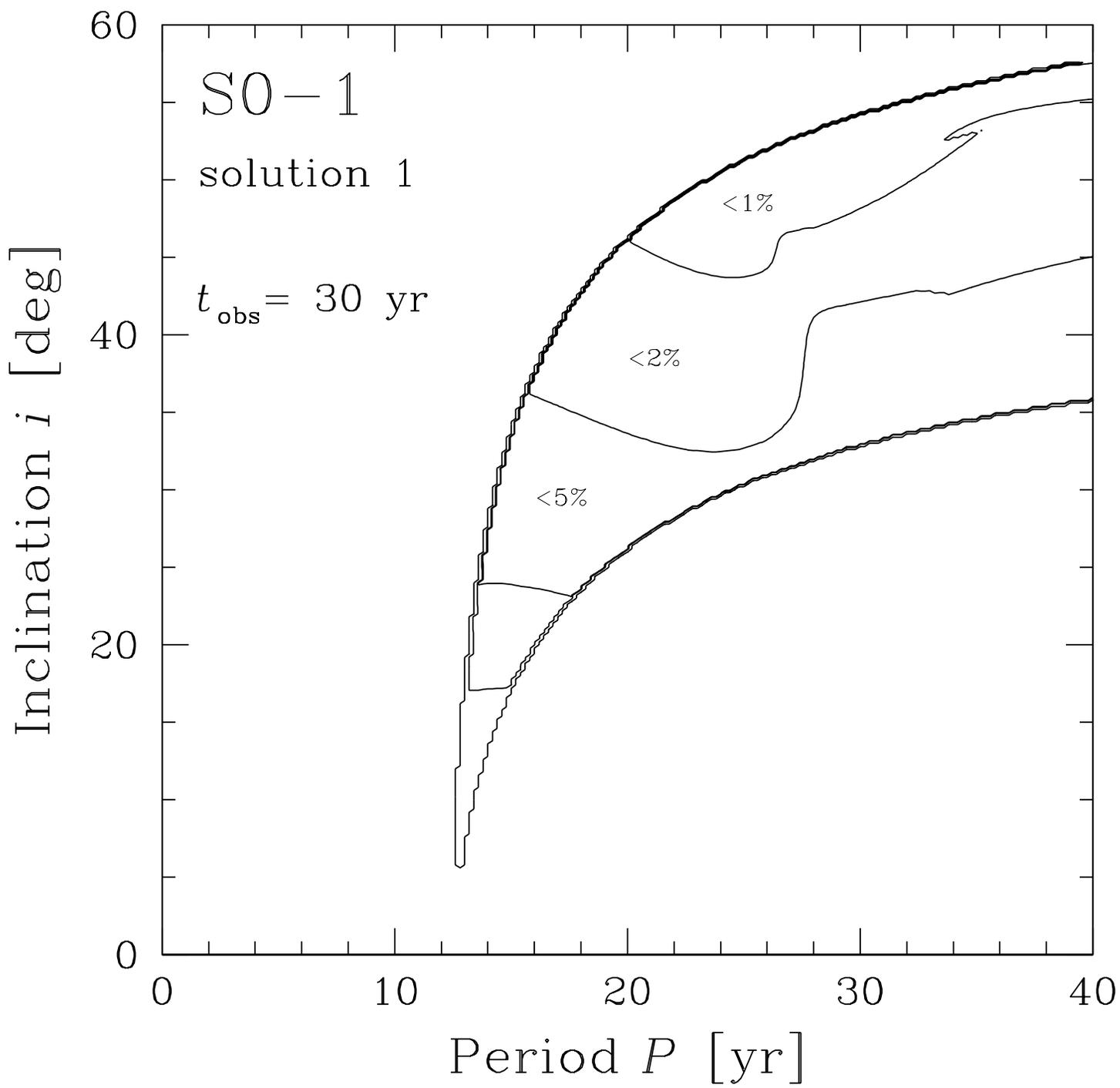
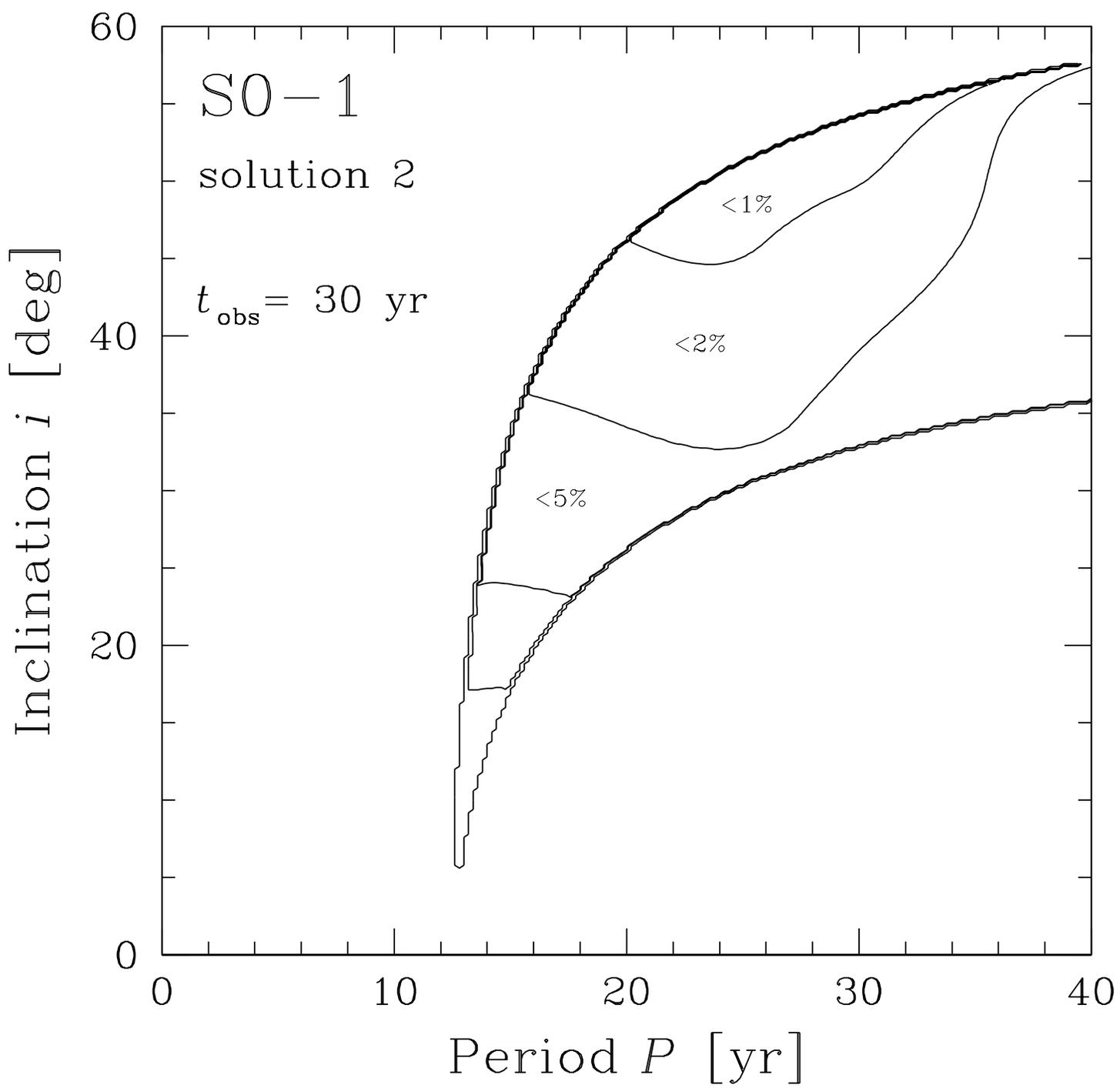
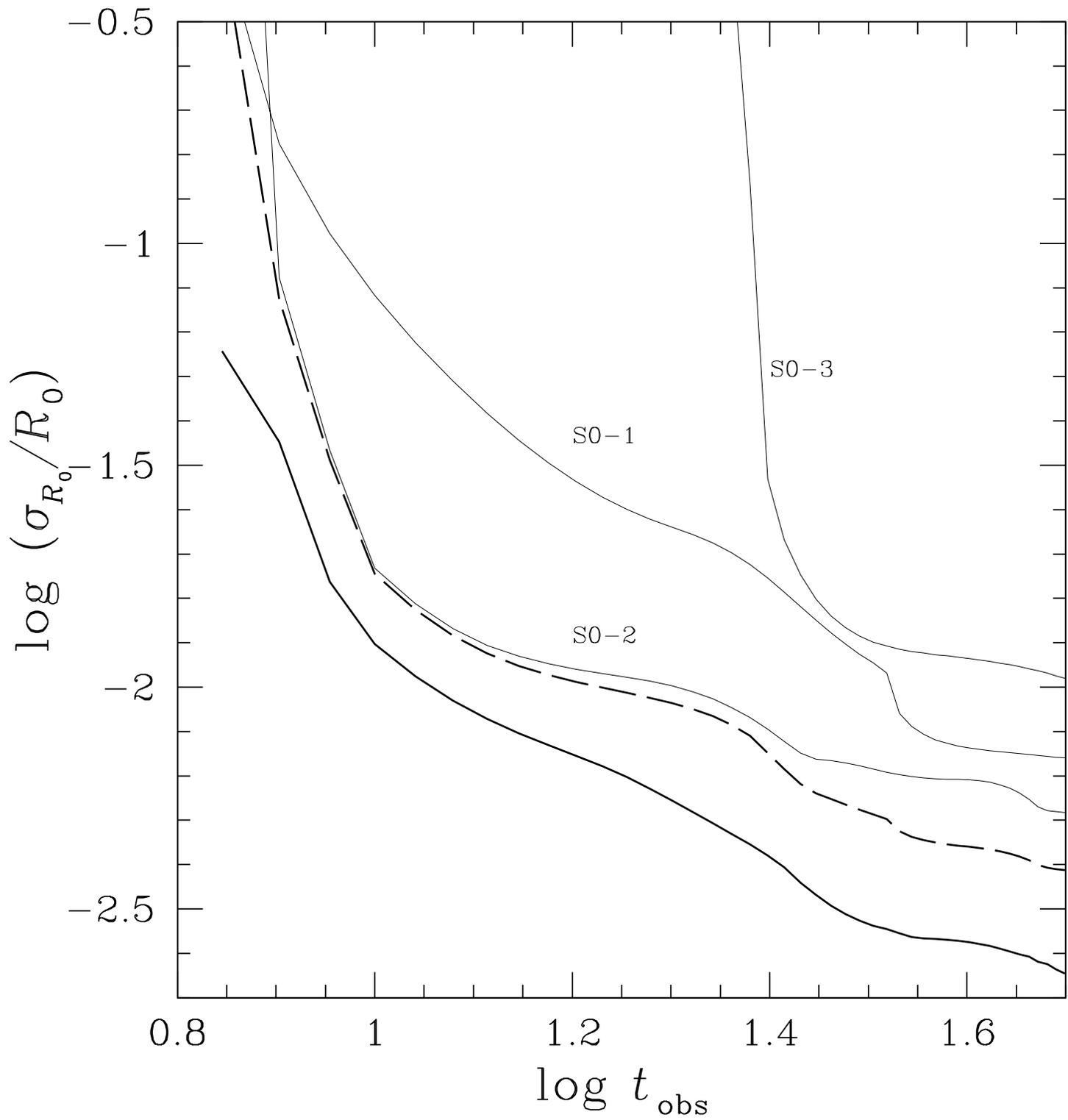
\end{document}